\documentclass[twoside,english,12pt]{cern2}

\usepackage{amssymb}

\usepackage{graphicx}
\usepackage{babel}
\usepackage{multicol}
\usepackage{setspace}

\newcommand{\registered}{\circledR}

\date{12 October 2006}

\begin{document}


\title{Field Emission Dark Current of \\ Technical Metallic Electrodes}
\author{F. Le Pimpec\thanks{Frederic.le.pimpec@psi.ch}, R.Ganter, R. Betemps\\
Paul Scherrer Institut \\ 5232 Villigen
\\Switzerland} \maketitle
\thispagestyle{empty} 

\thispagestyle{headings} \markright{SLS-TME-TA-2006-293}

\begin{abstract}

In the framework of the Low Emittance Gun (LEG) project, high
gradient acceleration of a low emittance electron beam will be
necessary. In order to achieve this acceleration, a -500~kV,
250~ns FWHM, pulse will be applied between two electrodes. Those
electrodes should sustain the pulsed field without arcing, must
not outgas and  must not emit electrons. Ion back bombardment, and
dark current will be damaging to the electron source as well as
for the low emittance beam. Electrodes of commercially available
OFE copper, aluminium, stainless steel, titanium and molybdenum
were tested, following different procedures including plasma glow
discharge cleaning.

\begin{verbatim}
    \PACS 29.25.BX \sep 52.80.Vp \sep 79.70.+q
 \end{verbatim}

\end{abstract}

\section{Introduction}

In the framework of the Low Emittance Gun (LEG) project, an X-ray
free-electron laser (FEL) based on a field emitting cathode is
expected to deliver six orders of magnitude higher peak brightness
than current state-of-the-art light sources, together with a
thousand times shorter pulses \cite{Bakker:FEL05,Candel}.

\medskip
To rapidly accelerate the electrons emitted by the electron
source, and keep the emittance low, a stable pulsed voltage in the
megavolt range is needed. The first project phase is to design and
test an ultra high vacuum (UHV) 500~kV pulser using a resonant
air-core transformer (Tesla coil) \cite{Paraliev:USA05}. A pulse
of 250~ns (full width at half maximum), -500~kV, working at 10Hz,
will be applied between the cathode holder and an extracting
anode. During this time, electrode materials should sustain the
field without arcing and the dark current should be kept as low as
possible. This dark current will ionize the residual gas as well
as desorbing neutrals and ions by the known electron stimulated
desorption process (ESD) \cite{redhead:97,yates:91}. Those ions
will be accelerated toward the cathode and the field emitter array
(FEA), in other words the electron source, and induce sputtering.
It is known that some surfaces are more sensitive than others to
very low energy ion bombardment. Measurable damage can already
occur at 500~eV \cite{Vossen:78,Cernusca,Cernusca:2003}. The
damage induced, from any kind of energetic ions, will then reduce
the electron emission and the lifetime of the field emitter, for
example in GaAs photocathodes, used as polarized electron sources
for accelerators \cite{Deicas:EPAC92,Grames:PESP04}. During the
after pulse, the reversed field will accelerate the ions toward
the extracting anodes producing a current of electrons which will
also back bombard the field emitter. The gas desorbed can induce a
pressure rise which might not disappear before the next pulse. In
the worst case, a plasma forms followed by breakdown and
sputtering of the anode material to the field emitter cathode.
Recent work, and thorough review of over a century of vacuum
breakdown research in many areas, \cite{llaurent,Werner,Goebel:05}
is still not sufficient to select electrode materials without
application specific testing.

In order to investigate the electrodes material, a DC high
gradient test stand was built to test different metals, which were
potentially suitable as electrodes in the pulser. The goal is to
find the most suitable material, for our needs, which can sustain
high field without breakdown, and emits almost no electrons. In
situ cleaning by plasma glow discharge was also tested to see
whether an improvement was noticeable in the mitigation of the
dark current. This technique of gas conditioning to lower the
field enhancement factor $\beta$ has already been reported
\cite{Noer:1982,Kobayashi:94,Kobayashi:97} and used successfully
in accelerators to process niobium accelerating cavities, see
description in \cite{Werner}, as well as for curing other issues
\cite{calder:1977}.

\section{System setup and electrode preparation}

\subsection{System setup}

\begin{figure}[htbp]
\centering
\includegraphics[width=0.8\textwidth, clip=]{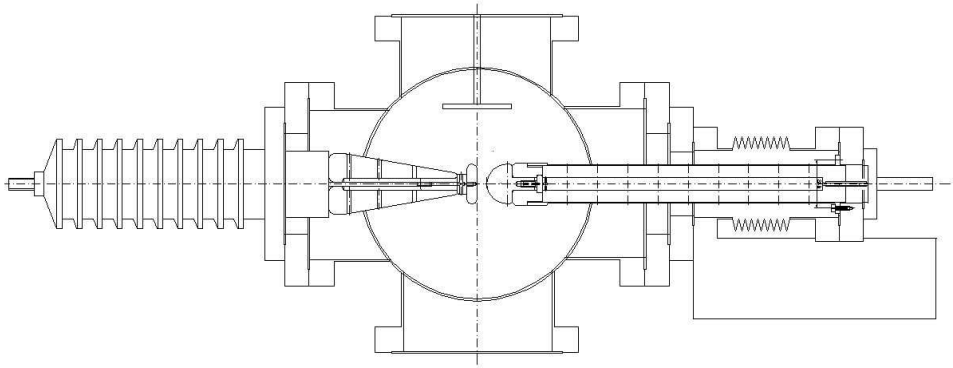}
\caption{Dark Current -100~kV DC test stand.} \label{figTestStand}
\end{figure}

\begin{figure}[hbp]
\centering
\includegraphics[width=0.8\textwidth, clip=]{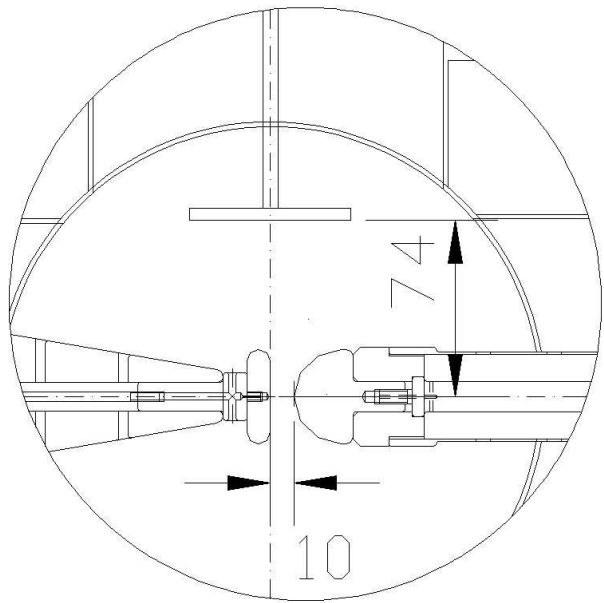}
\caption{Dark Current -100~kV DC test stand. Zoom over the area
containing the cathode (flat electrode) the anode (mushroom) and
the plasma electrode located at 7~cm from the center of the
system} \label{figTestStandzoom}
\end{figure}

The ultra-high vacuum (UHV) system shown in
Fig.\ref{figTestStand}, outside its metal confinement bunker for
radiological protection, is pumped by a 150~l/s diode ion pump.
The average pressure reached is in the low 10$^{-9}$~Torr scale
after a quick bake of the ion pump. A more thorough bake brings
the pressure down to the mid 10$^{-10}$~Torr. An injection line, a
leak valve and a Torr capacitance gauge allow the controlled
injection of a chosen gas into the system to prepare the glow
discharge between the electrodes. In order to clean the anode and
the cathode at the same time, a third electrode polarized
positively in respect to the test electrodes is used,
Fig.\ref{figTestStandzoom}. The distance of this third electrode
from the center of the chamber is a constant.

A negative and continuous 0 to 100~kV bias is applied to the
cathode through an insulating ceramic, (on the left in
Fig.\ref{figTestStand}). The anode is grounded. The capacitance of
the system is close to 300~pF, hence the potential energy stored
is 1.5~J at 100~kV. The current flowing from the cathode to the
anode is measured across a 1~M$\Omega$ resistor with a digital
Fluke{\texttrademark} voltmeter. The gap separation between the
electrodes is adjustable via a translation feedthrough and is
controlled with a mechanical comparator. The sagging due to the
weight at the end of the rods means that the two electrodes are
not centered on each other. We do not expect this off centering to
be of any consequence to the high voltage processing. However, it
explains the off center damages seen on cathodes. Accurate
measurement of the current to the 50~pA level is achieved.

\subsection{Electrode choice}

Given the long history of research on vacuum breakdown, it seems
that the choice of the electrodes should be easy. However, as
there is no universal quantifying theory to explain the process of
vacuum breakdown depending on the material, and its surface state
(physical and chemical), it is necessary to do our own testing for
our own application. In order to pick the most appropriate
material, we study some elemental properties. Many tables of
elements are then compiled in order to make an informed choice. In
our case, the electrodes should sustain a DC pulse of 500~kV and
not produce or have a low electron dark current. Also a FEA will
be installed in the middle of the cathode. If any arcing occurs,
sputtered, or vaporized, anode material will deposit on the FEA.
This can lead to destruction of the FEA, and the necessity to
exchange it; in our system this is a complex and time-consuming
procedure. Table.\ref{tabElectrchoice}, compares the secondary
electron yield (SEY) the sputtering rate, the melting point, and
the tensile modulus of the candidate elements.

From Table.\ref{tabElectrchoice}, Cu and Au appear to be bad
candidates. Results obtained with RF waveguide support this point
\cite{dolgashev:EPAC02}. Despite its good electronic and ionic
properties Al should be discarded, as the combination of melting
point and elastic modulus is low compared to other materials. Al
will probably coat a FEA thoroughly in case of arcing, as we also
found with a gold coated anode. The FEA coating problem further
implies that spark processing to reach high gradient, despite
being efficient, should be avoided
\cite{Williams:72,Kildemo:2004}, unless it is possible to protect
the FEA. Some of the other materials which look good in this
table, can probably be discarded, due to their yield strength
versus the temperature or their electrical or thermal
conductivity. Also the choice of the cathode and anode should be
made separately as a good cathode material might not be so well
suited as an anode.  The respective choices depend on the geometry
of the system, and the removal of any heat generated by the dark
current.

\subsection{Electrode preparation and testing}

\begin{figure}[htbp]
\begin{minipage}[t]{.5\linewidth}
\centering
\includegraphics[width=0.95\textwidth,clip=]{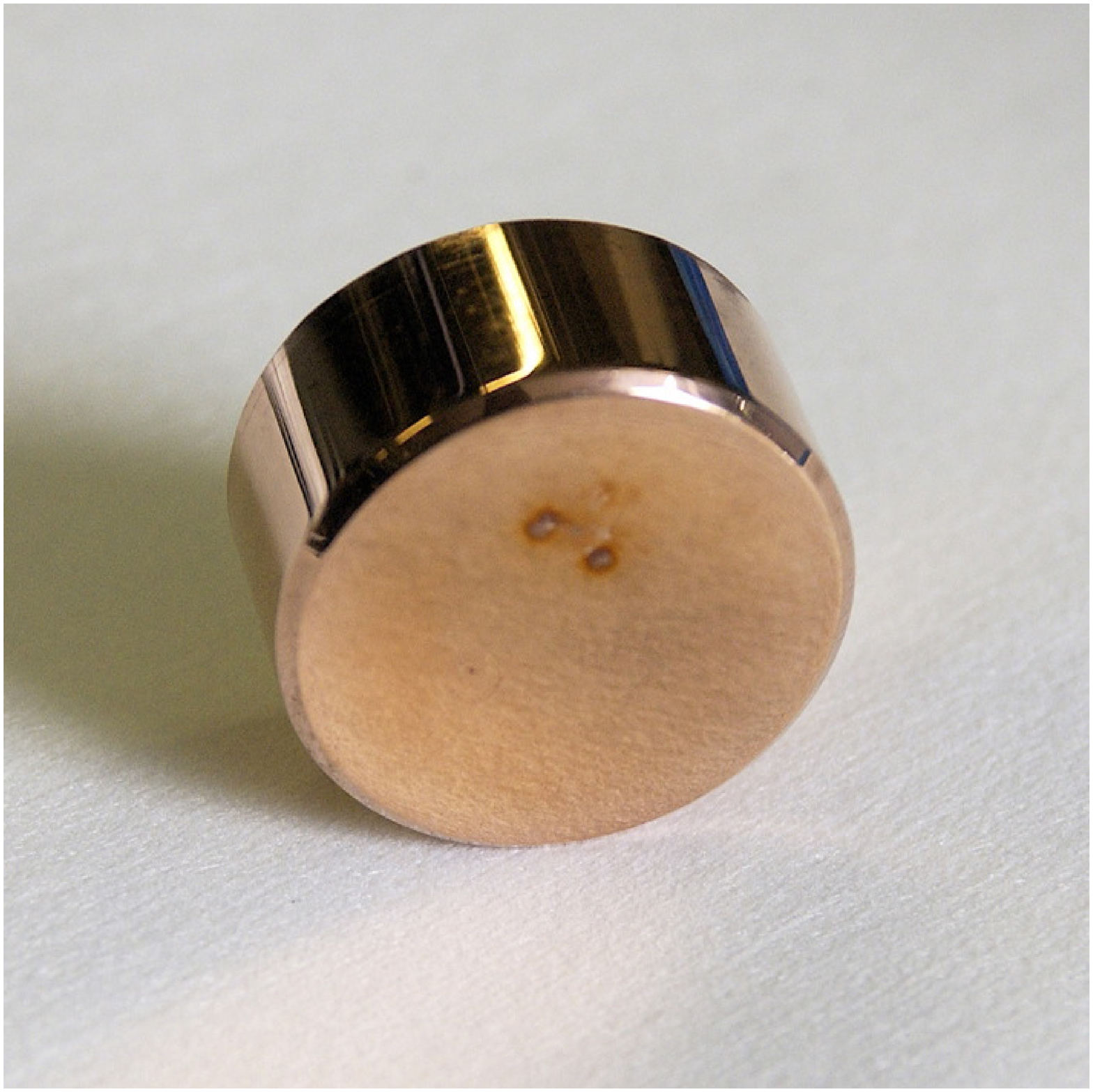}
\end{minipage}%
\begin{minipage}[t]{.5\linewidth}
\centering
\includegraphics[width=0.92\textwidth,clip=]{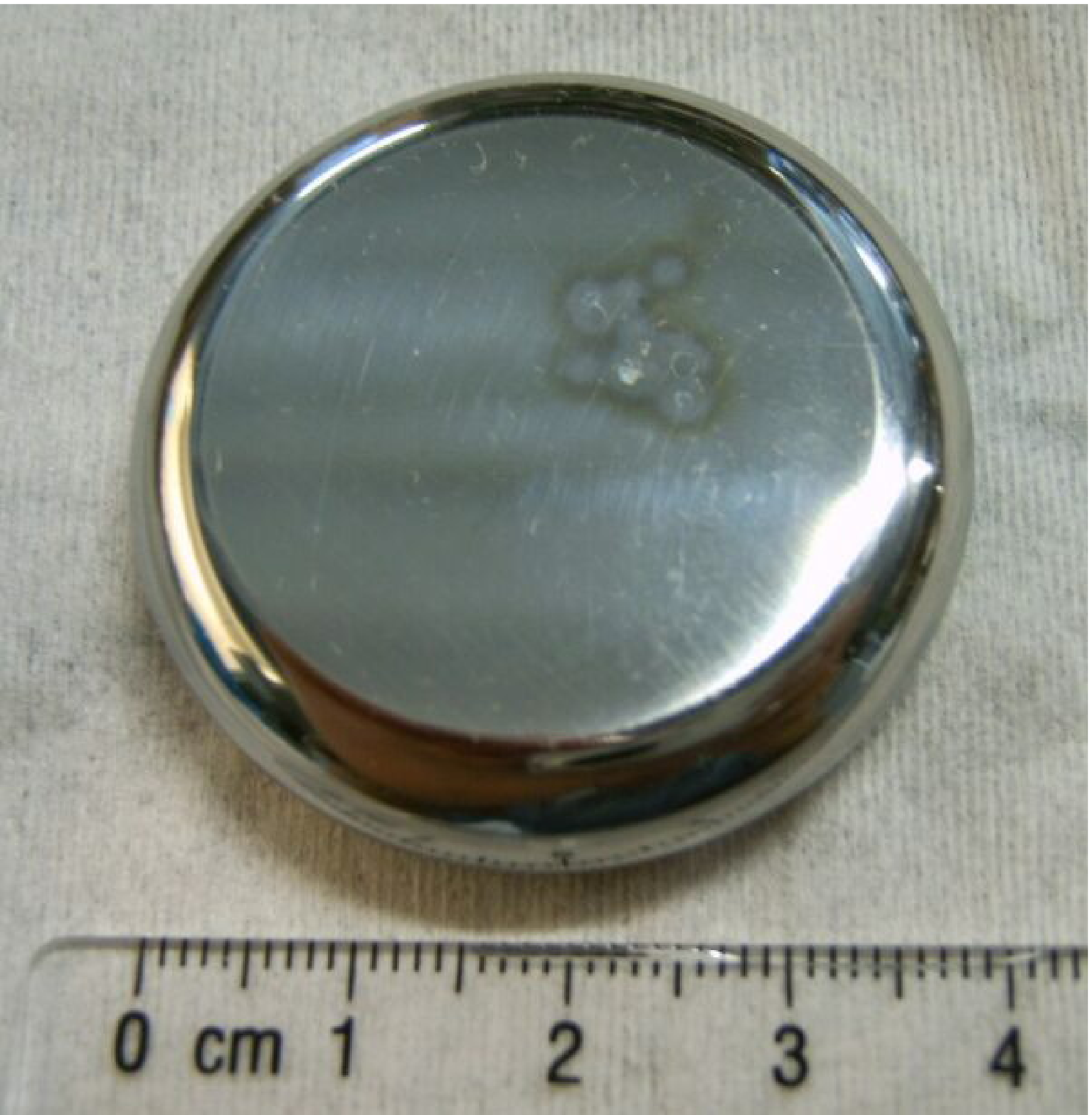}
\end{minipage}
\caption{Cleaned Cu cathode after high voltage testing, left.
Damaged Mo cathode, right picture. Damage can be clearly seen.}
\label{figCuMocathode}
\end{figure}

\begin{figure}[htbp]
\centering
\includegraphics[width=0.9\textwidth, clip=]{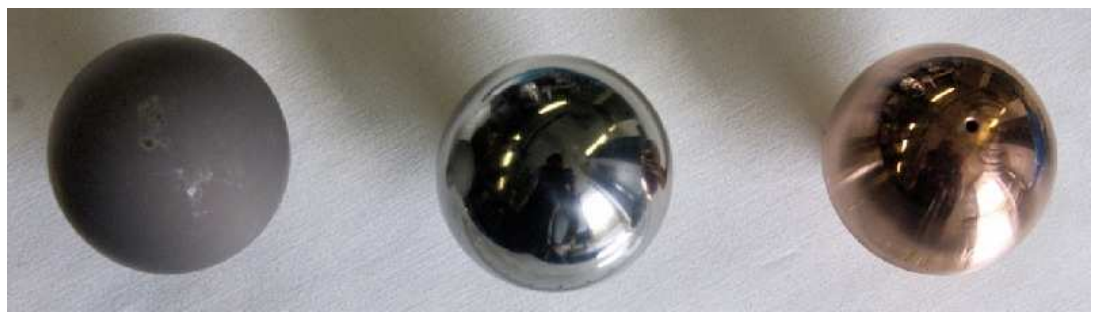}
\caption{Three anodes used for HV testing, Ti on the left(grey
black colour, after vacuum firing), Mo in the center and Cu on the
right. Pictures taken after HV testing. Damage can be seen on the
Ti electrodes} \label{figTiMoCu}
\end{figure}

All our electrodes tested have the same shape, see
Fig.\ref{figCuMocathode} \& \ref{figTiMoCu} for the cathode and
anode, respectively. The mean roughness was, by design, defined to
be less than R$_a$ $\leq 0.2 \mu$m. In Fig.\ref{figTiMoCu}, the
copper anode, has a hole in the middle. This hole was made to
mimic the behaviour of the extracting anode of the 500~kV pulser.
None of the other electrodes have this hole. The R$_a$ of the
electrodes was checked after high gradient testing.

The electrodes were cleaned using acetone and alcohol in an
ultra-sonic bath, before installation in the UHV system. This is
the "as received" state. Unless specified otherwise none of the
surfaces have been mirror finished. All materials were
commercially obtained from Goodfellow{\texttrademark}. Technical
materials refer to commercially available material, which is
exposed to air before installation.

Ti and Mo electrodes were fabricated by the same machining
company. Electrodes were thoroughly cleaned in acetone and alcohol
before use. Ti electrodes were installed after cleaning and
tested, and Mo electrodes were vacuum fired at 900$^{\circ}$C for
3~h. After Mo testing, Ti electrodes were also vacuum fired and
reused. From the literature, it was shown that heating up the
material is beneficial in improving the breakdown strength
\cite{diamond1:98}.

The processing histories of the materials tested are summarized in
Table.\ref{tabHistoryelectrode}. The procedure of high gradient
conditioning is the same for all the cathodes. The voltage between
the electrodes is applied for a given gap, 4~mm, 3~mm, 2~mm,
1.5~mm and then 1~mm. The voltage is raised slowly, waiting for
stable conditions, by discrete steps, to 60~kV and then the gap is
diminished with a reduced voltage equal to the previously obtained
static electrical field. It was found that above 70~kV arcing
sometimes happened somewhere else in the system.

During conditioning, soft breakdowns might occur. During those
breakdowns, current is measured and the pressure can increase by a
factor 10. When observed, the voltage is, usually, manually
reduced. The pressure recovers in a minute or two, and the voltage
is again raised slowly to the previous level.

In this study, we do not reproduce quality preparation achieved in
\cite{Furuta:Nima2005}. Instead, our aims were to see what is the
behaviour of a technical material prepared using a less stringent
procedure. Once the electrodes are installed under vacuum, our
intention is to use plasma glow discharge (PGD), known to be an
efficient way of cleaning surfaces and removing the contamination
which can promote field emission. We do not think the exceptional
surface preparation achieved by the procedure in
\cite{Furuta:Nima2005} will survive an aggressive PGD.

Plasma glow discharge was usually applied after we reached 1~nA of
current at 1~mm gap from the as received state. The gases injected
for the PGD are usually a mixture of He and Ar, with a composition
of 50\% He and 50\% Ar. Sometime pure Ar is used. The total
pressure range was between $\sim$0.15~Torr and $\sim$0.25~Torr.
Noble gases are chosen to avoid chemically attacking the surfaces.
Helium is chosen because, for the same energy, its sputtering
potency is less than Ar. The gases come directly from a compressed
gas cylinder and are injected via a leak valve. A +400~V to +600~V
bias is applied between a third electrode,
Fig.\ref{figTestStandzoom}, and the two electrodes to be tested.
The pressure and the energy of the ions in the PGD are adjusted,
so that the plasma wraps around the electrodes. The distance
between the three electrodes are around 6~cm and the time of the
PGD can last between 40 to 60~minutes.

Finally, as it is known from literature that pressure can affect
the breakdown onset threshold, and that dark current appearance is
affected by the gas species \cite{llaurent}, the system is baked
not only after air venting but also after each plasma. By this
means we minimize any role that the pressure and the gas
composition would have in FE or arcing.

\section{Results}

Before presenting results obtained with our electrodes, it is of
importance to bear in mind results obtained by Furuta \emph{et al}
\cite{Furuta:Nima2005,suzuki:2001} and Diamond
\cite{diamond1:98,diamond2:98}. According to Furuta's publication,
the design of their electrodes are equivalent to ours. They have
obtained, for stainless steel, Cu, Ti and Mo with mirror finished
surfaces, the results summarized in Table.\ref{tabFuruta}. Those
results have been obtained not only with mirror finished surfaces,
but the assembly of their system and the mounting of their
electrodes, were done in class 1 and class 10 clean room
condition, respectively.

All the current vs electric field plots presented in
\cite{Furuta:Nima2005} and in this work can be fitted using $I = c
\ E^2 \ e^{-a/E}$, see Fowler-Nordheim equation (\ref{EquFN}),
with I current, c and a constants, and E applied electric field.
From those fits F-N parameters, area and the field enhancement
$\beta$, can be extracted.

\begin{equation}
I = A \cdot \frac{1.5 \cdot 10^{-6}}{\Phi} E_s^2 \cdot
e^{\frac{10.4}{\sqrt{\Phi}}} \cdot exp \ ({\frac{-6.83 \cdot 10^7
\Phi^{\frac{3}{2}}}{E_s}}) \label{EquFN}
\end{equation}

where $E_s = \beta \cdot E$ and the work function, for Mo, $\Phi$
is taken equal to 4.2 eV. However, we will not go further into the
comparison between our measurements and theory, because the goal
of this paper is to report on practical surface conditioning
procedures to achieve stable operation under a high electric
field. It is also to report on the erosion of the materials upon
the field processing.

\subsection{Aluminium Results}

\subsubsection{Al-Al electrodes}
\begin{figure}[tbp]
\centering
\includegraphics[scale=.5,width=0.6\textwidth,angle=-90, clip=]{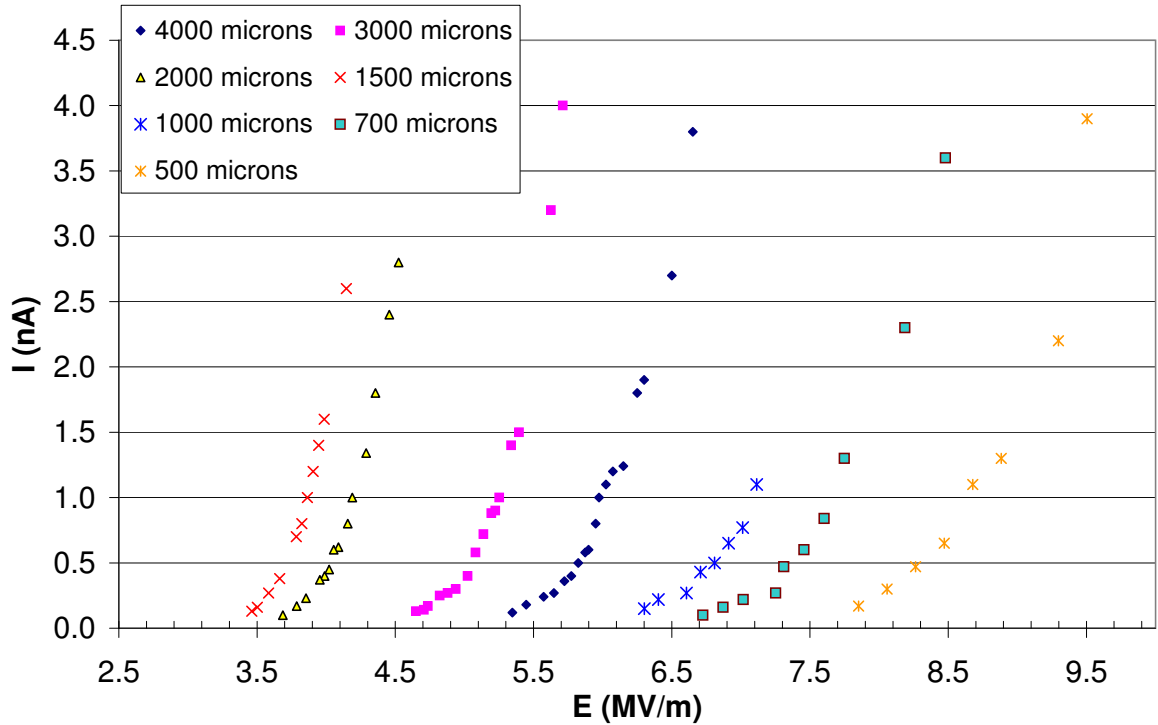}
\caption{Dark current of Al electrodes after a system bake at
190$^\circ$C for 100h and after electrical conditioning of 29~kV
at a 4~mm gap drawing 7~nA of current from the cathode to the
anode.} \label{figdrkcurrentAl}
\end{figure}

Pristine, as received Al electrodes were tested. The first test
after a thorough bake of the chamber led to a dark current of 1nA
at a gap of 1 mm for a field of 7.1~MV/m. The gap between the
electrodes was set at 4~mm and the electrodes were conditioned
overnight by applying 29~kV and drawing 7~nA of current. The next
day the dark current increased to 13~nA. Several dark current
curves were then produced and compared to the as received test,
Fig.\ref{figdrkcurrentAl}. The obvious conclusion is that our DC
electrical conditioning did not lead to any improvement.

\begin{figure}[tbp]
\centering
\includegraphics[scale=.5,width=0.6\textwidth,angle=-90, clip=]{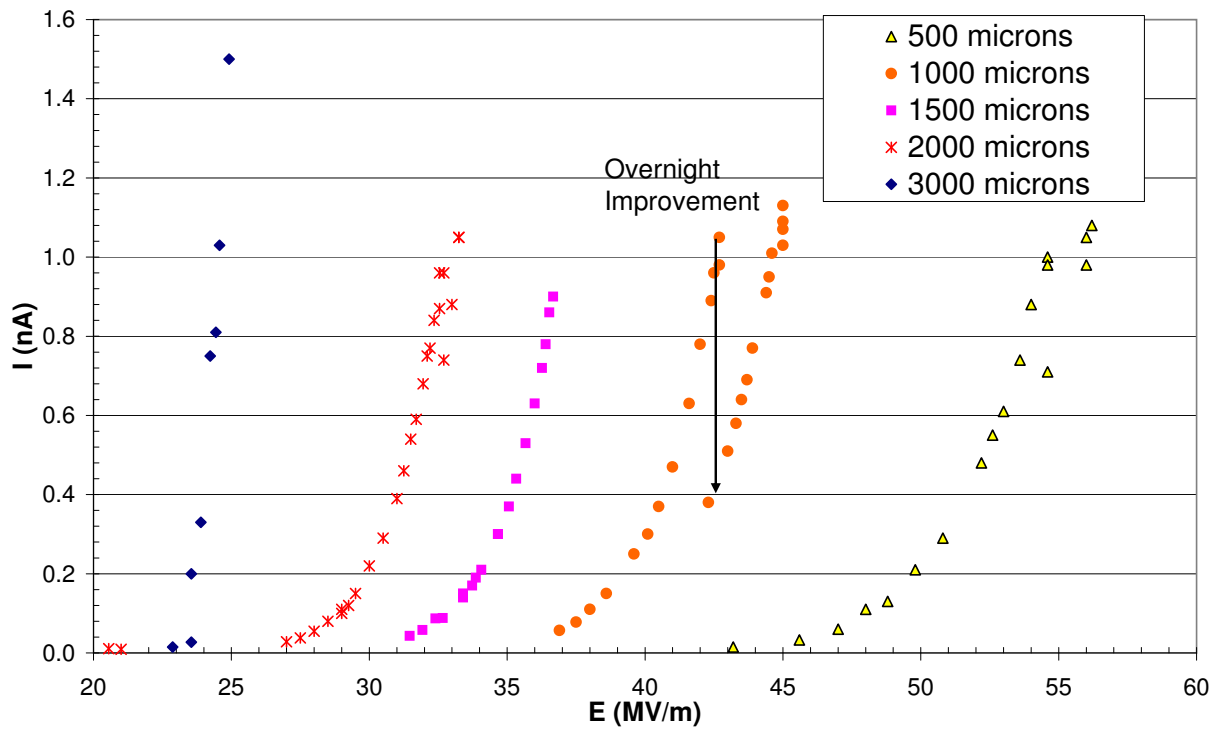}
\caption{Dark current of Al electrodes after an Ar Glow Discharge,
following the previous He processing and high field conditioning.}
\label{figAlArGD}
\end{figure}

We next tested cleaning and conditioning using an He plasma of
0.26~Torr. The sputtering rate of 500~eV He ions on Al is 0.16
\cite{Vossen:78}. The 1~nA at 1~mm gap was reached for a field of
13.5~MV/m. Subsequent He plasma continued to improve the results.
However, a few breakdowns occurred during the voltage processing.
An ArGD at a pressure of $\sim$0.1~Torr produced the best results
which are presented in Fig.\ref{figAlArGD}. The 1~nA at 1~mm gap
(full circles) was reached for a field of 42~MV/m. When leaving
the system at this level of field and dark current further
improvement, decrease of dark current over time, is observed
Fig.\ref{figAlArGD} full circles. In subsequent tests, also with
different materials, this improvement was occasionally observed.
However, in some cases the current increased to more than the
double the previous value.

In the next test an He-Ar plasma was used to clean the same
electrodes. The behaviour, after pumping out the noble gases, was
that no FE was observed until breakdown. The Al electrodes held
stably (12hours) a field of 42~MV/m without dark current at 1.5~mm
gap. Whilst the field was at 45~MV/m an arc occurred and an
emission current of 350~nA could be measured. Nevertheless, damage
due to this and further breakdowns during the voltage processing,
were not severe; the current at 1~mm, for a field of 41.5~MV/m was
still only 1~nA. Finally, the best result obtained by He-Ar GD,
was a field of 52~MV/m at 1~mm without dark current. However, at
some point an arc more violent than the previous ones, damaged the
cathode so that no more GD was able to restore the holding of the
high electric field. A summary of the performance obtained with Al
electrodes is shown in Table.\ref{tabAlRslt}.


\subsubsection{Mirror Finished Al Cathode}
\begin{figure}[htbp]
\begin{minipage}[t]{.5\linewidth}
\centering
\includegraphics[width=0.95\textwidth,clip=]{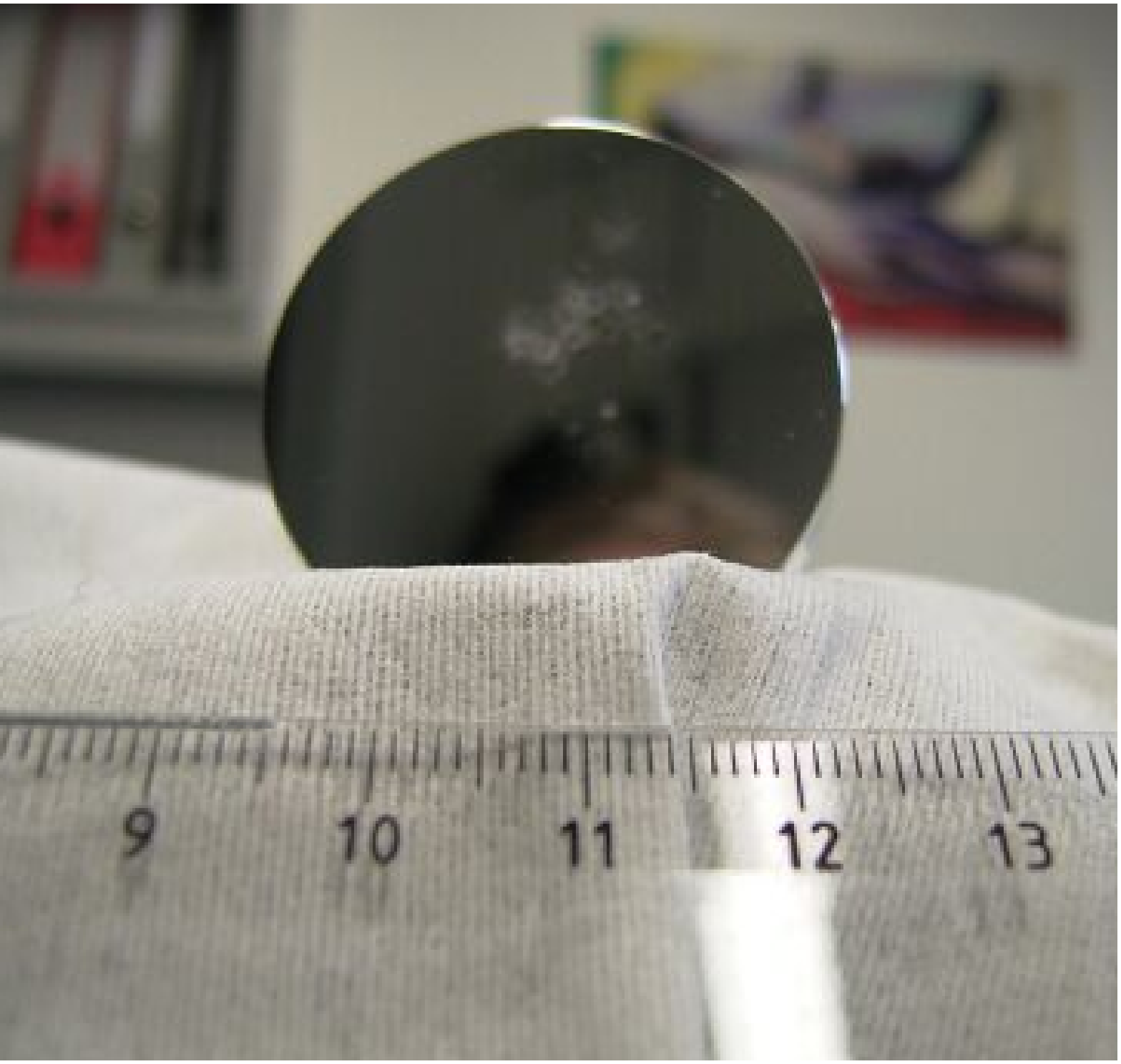}
\end{minipage}%
\begin{minipage}[t]{.5\linewidth}
\centering
\includegraphics[width=0.95\textwidth,clip=]{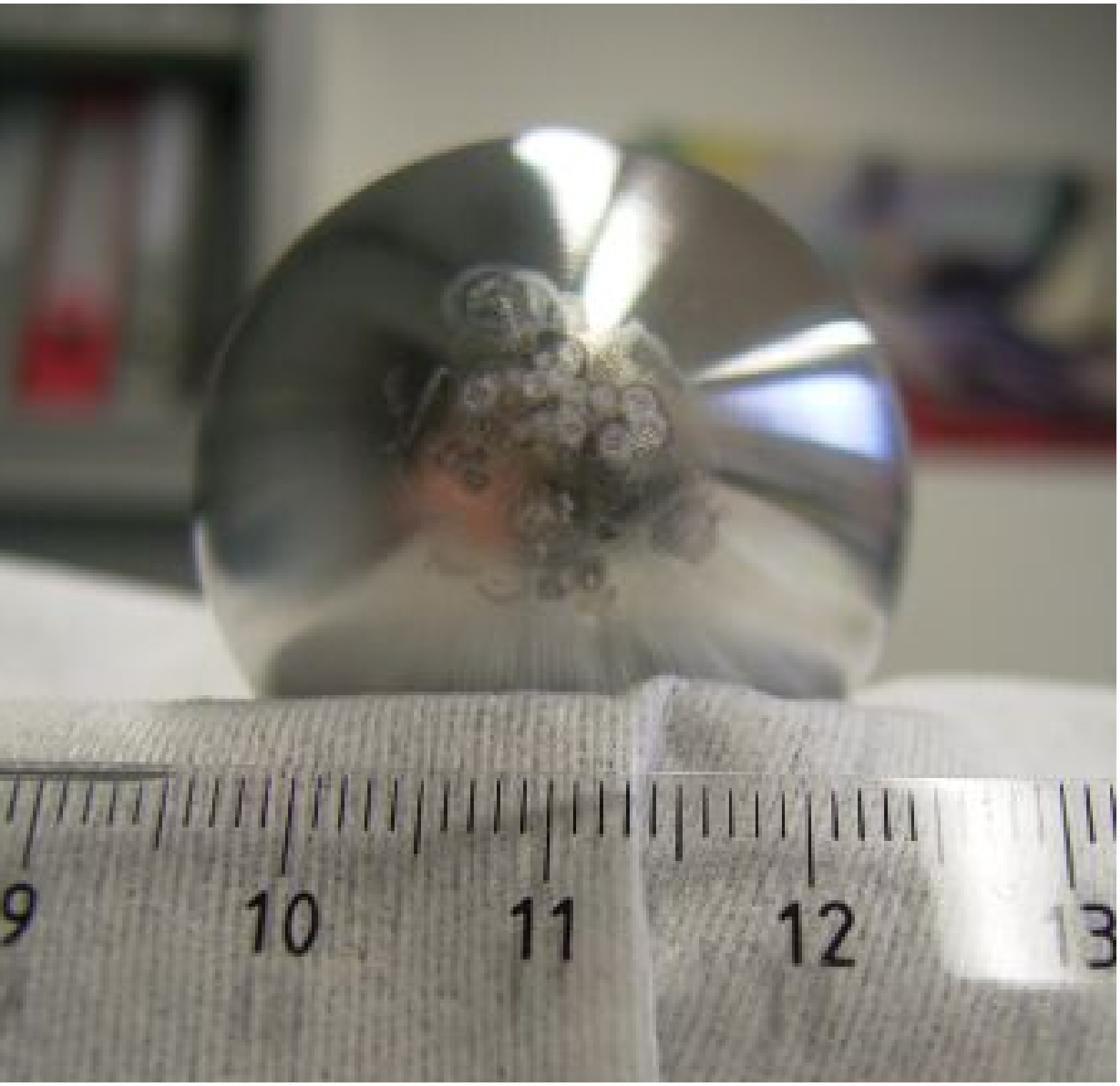}
\end{minipage}
\caption{Breakdown damage withstood by the Al cathode (mirror
finished-left) and anode (right) during the conditioning period.
Scale in cm.} \label{figAlelectrodedmge}
\end{figure}

We then replaced the damaged Al cathode by a pristine mirror like
finish Al (6082) cathode, machined to an R$_a$ of 3~nm. The
previously damaged mushroom anode was reused, after wiping with
alcohol before reinstallation in the UHV system. The damage on the
anode was localized around the summit of the anode, and resembled
the right-hand picture in Fig.\ref{figAlelectrodedmge}.

With this configuration, the as received system held stably
without dark current, up to an electric field of less than
36~MV/m. At this value, an arc occurred. Inspection of the cathode
through the viewport of the UHV system showed pitting damage. An
He-Ar GD was then applied to cure and clean the electrodes.
Results are summarized in Table.\ref{tabAlRslt}. The column
labelled $<$0.05~nA shows the field strength held without
measuring any FE. The system held 90~MV/m electric field at
750~$\mu$m, and broke down at 92~MV/m. The vacuum arcing was so
severe that no further PGD was able to restore such fields. Final
damage to the Al electrodes is shown in
Fig.\ref{figAlelectrodedmge}. The high field sustained shows that
the breakdown is cathode initiated as the cathode was pristine and
the anode was already severely damaged.

\subsection{Copper results}

Oxidized copper electrodes were tested solely after a PGD. Even
after a PGD and the voltage processing, the electrodes are still
very oxidized. Cleaner spots around the hole of the anode were
observed at the end of the testing. Damage on the cathode was also
visible. The electrodes were then cleaned by chemical etching by
use of a phosphoric acid based solution Polynox$^{\registered}$.
They were subsequently rinsed with tap water and after drying,
cleaned with ethanol. A last test was conducted by installing a
mirror diamond turned OFHC Cu cathode (R$_a \sim 3$~nm) and the
already used Mo and then a SS anode.

\subsubsection{Cu-Cu electrodes}

The results obtained seem to show that there is no influence from
the anode hole in the achievement of the high gradient, as our
results are similar to what was obtained in
\cite{Furuta:Nima2005}, see Table.\ref{tabFuruta} for clean
copper.

In comparison to Al electrodes, craters in the Cu were neither as
deep, nor as extended : see the 2 spots in
Fig.\ref{figCuMocathode}, compared to damages in
Fig.\ref{figAlelectrodedmge}. All breakdown damage on the Cu anode
remained localized around the hole of the mushroom. That suggests
two possibilities without excluding a combination of the two. The
energy in the arc was not sufficient to vaporize the Cu materials,
by melting and sputtering the melted Cu, despite the fact that Cu
has a higher sputtering rate than Al. Or the field was not strong
enough to pull out droplets of Cu which could have then been
vaporized \cite{Mesyats:ICPIG03}.

\subsubsection{Mirror finished Cu cathode}
As results from Al seemed to indicate that the breakdown is
cathode initiated, a combination of a mirror finished cathode with
the previously used Mo (vacuum fired ) and stainless steel(SS)
anodes was tested. Those anodes had sustained damaged far less
significant than the Al anode pictured in
Fig.\ref{figAlelectrodedmge}. Results of the Cu-Mo conditioning
are shown in Table.\ref{tabCuRslt}. The conditioning of the "as
received" electrode followed the "breakdown processing" or spark
processing scheme, until a more severe breakdown brought the Cu
cathode to emit at the level of 800~nA, at 3~mm gap. During the
first 500~eV ArGD, small breakdowns could be seen on the Cu
cathode, which were probably dust burning away. Results obtained
after plasma processing improved the situation but not to the
level of the two clean Cu electrodes. It was expected that higher
fields could be reached, as in the case of using two Cu
electrodes, Table.\ref{tabCuRslt}, or two Mo electrodes (see
Table.\ref{tabTiSSMorslt}).

The Mo anode was exchanged for a SS anode, and the Cu cathode was
turned 180$^\circ$ on its axis. Because the anode arm sags, damage
on the cathode is not localized on the center of the anode, hence
allowing a pristine area to be exposed. The maximum field held
with this pair was 11~MV/m at 3~mm gap, after an ArGD. From these
last results, no conclusion should be drawn on the coupling of
this last pair as the Cu cathode was already exhibiting an oxide
color. The presence of the oxide can be due to oxygen present in
the ArGD, or from oxygen being released from the Mo anode during
high voltage conditioning see \S \ref{xpsanalysis}. The copper
oxide is present from the center of the Cu sample to 5~mm from the
sample circular edge. Microscopy and surface analysis did not
reveal any material transfer from the anode (Mo or SS) to the Cu
cathode.

\subsection{Stainless steel, titanium and molybdenum results}

Main results obtained, at 1~mm, for stainless steel (SS), Ti and
Mo are summarized in Table.\ref{tabTiSSMorslt}.

\subsubsection{Stainless Steel}
SS electrodes were electrically processed the same way as Al
electrodes. After plasma treatment, the best field achieved was
68~MV/m with dark current below the 10~pA detection limit. The
final breakdown, was not recoverable by the use of a PGD. Upon
removal of the electrodes, damage was located at the top of the
mushroom. However, craters were not as deep or as wide as for Al
electrodes (in Fig.\ref{figAlelectrodedmge}). The damage is less
extended than that observed on the Cu electrodes. If we look at
the data in Table.\ref{tabElectrchoice}, this is not surprising.
Effectively, both the melting point and the Young's modulus of SS
are far above copper values. However, as tiny amounts of vaporized
or sputtered material from the anode can be prejudicial to the
working of FEAs, avoiding even soft breakdowns seems to be a good
strategy.

\subsubsection{Titanium}
\begin{figure}[hbp]
\centering
\includegraphics[scale=.5,width=0.6\textwidth,angle=-90, clip=]{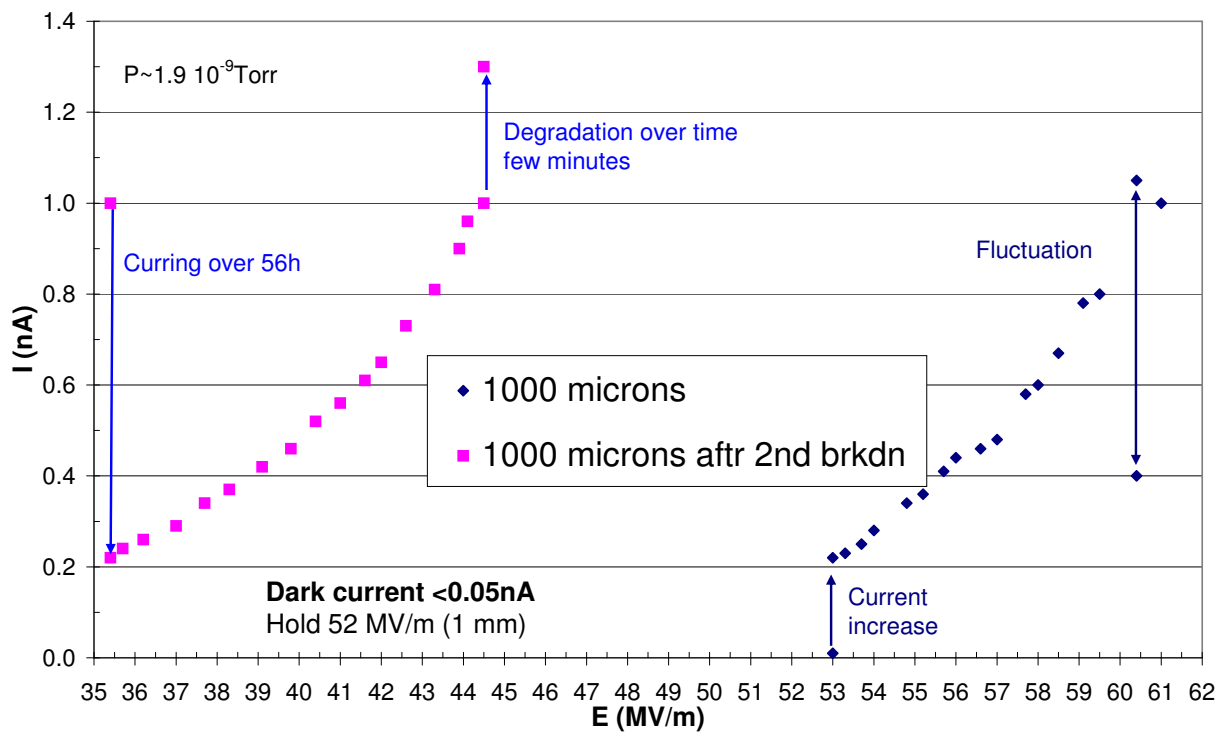}
\caption{Dark current evolution of Ti electrodes after a 2$^{nd}$
Ar Glow Discharge. The square curve is obtained after a surface
breakdown at 62~MV/m} \label{figTiArGD}
\end{figure}

In the case of Ti, we have observed after plasma treatment, a
stable field holding at 52~MV/m without FE. Above 53~MV/m, dark
current appeared and reached 1~nA at 62~MV/m, cf
Fig.\ref{figTiArGD} (diamonds). The field held at this value for a
few minutes until arcing, which brought the current above
1~$\mu$A. The field was then reduce to 35~MV/m in order to get a
1~nA current value, cf Fig.\ref{figTiArGD} (squares.) This current
value decreased over 56~hours of 35~MV/m field processing. A few
MV/m were then gained to bring back the current to 1~nA.

It is interesting to mention that after a soft breakdown leading
to dark current emission around 1~$\mu$A, sometime no rise in
pressure is detected. At a pressure of 2.10$^{-9}$~Torr, gas
released by ESD process can in principle be detected. A simple
calculation will show that such a current, 1~$\mu$A, will produce
a gas flux of $\sim$3.10$^{-9}$~Torr.l.s$^{-1}$, hence an increase
of pressure of $\sim$2.10$^{-11}$~Torr; assuming an ESD
coefficient of 10$^{-2}$~molecule/electron. This is below the
resolution of our instrumentation, and of most total pressure
gauges.

\begin{figure}[hbp]
\centering
\includegraphics[scale=.5,width=0.6\textwidth,angle=-90, clip=]{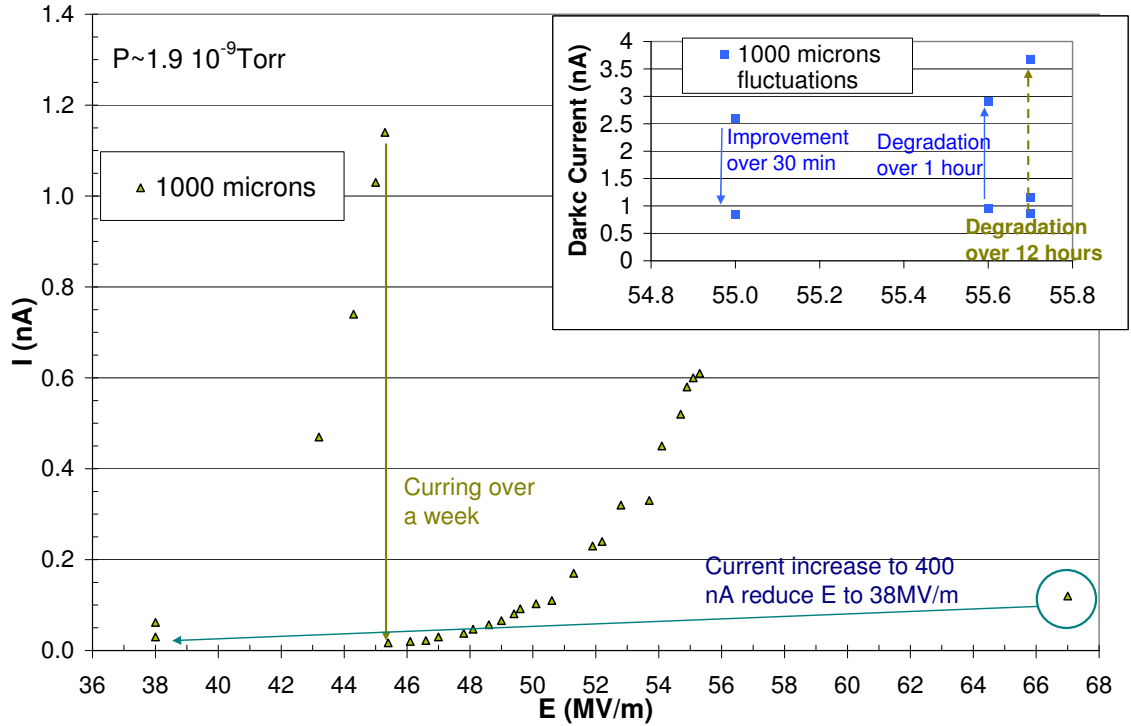}
\caption{Dark current evolution of Ti electrodes after a 3$^{rd}$
Ar Glow Discharge.} \label{figTiArGD3}
\end{figure}

A subsequent ArGD for 90~minutes at 580~eV under pressure varying
from 0.156~Torr to 0.174~Torr was performed. During this PGD, the
Ar gas was evacuated several times and then replaced by new Ar
coming from the cylinder. The field held after this plasma was
61~MV/m. The voltage was then increased and fluctuations in the
emitted current below 50~pA were observed. Between 63~MV/m and
67~MV/m the dark current was fluctuating around 0.1~nA, see
Fig.\ref{figTiArGD3}. Field conditioning over a week showed, as in
the case of Al or SS, an improvement in the dark current emission.
The final stable electric field obtained for 1~nA at 1~mm gap is
55~MV/m, see insert in Fig.\ref{figTiArGD3}. Above this level of
field, the dark current does not diminish but it increases with
time, Fig.\ref{figTiArGD3}.

Upon inspection of the electrodes after their extraction from the
chamber, damage spots can be seen on both electrodes. Damage is
located on the anode around the top of the mushroom, and 4
separate spots can be seen on the cathode spread over a trail of
1~cm length. Damage on the Ti anode is similar to that seen on the
SS and Cu anodes, in extension and morphology (melted area).

\subsubsection{Molybdenum}

Vacuum fired Mo electrodes were installed in the system, the best
results are summarized in Table.\ref{tabTiSSMorslt}. A simple 25Hz
optical camera was used to view the electrodes, hence the space in
between. When running, no light is present in the system
enclosure. During the soft events, a flash is seen on the TV
screen and a jump in current intensity, below 0.1~nA, is measured
sometimes associated with a pressure increase. The flash is
localized and takes the full space occupied between the
electrodes. During harder breakdowns, sound can be heard coming
from the enclosure. After these soft breakdowns, the voltage could
still be increased. In order to have a chance to detect breakdown
precursor, fast acquisition and very high sensitivity, to single
photon, are mandatory. However, such a simple optical system can
be used as an interlock protection for our FEL.

\begin{figure}[hbp]
\centering
\includegraphics[scale=.5,width=0.65\textwidth,angle=-90, clip=]{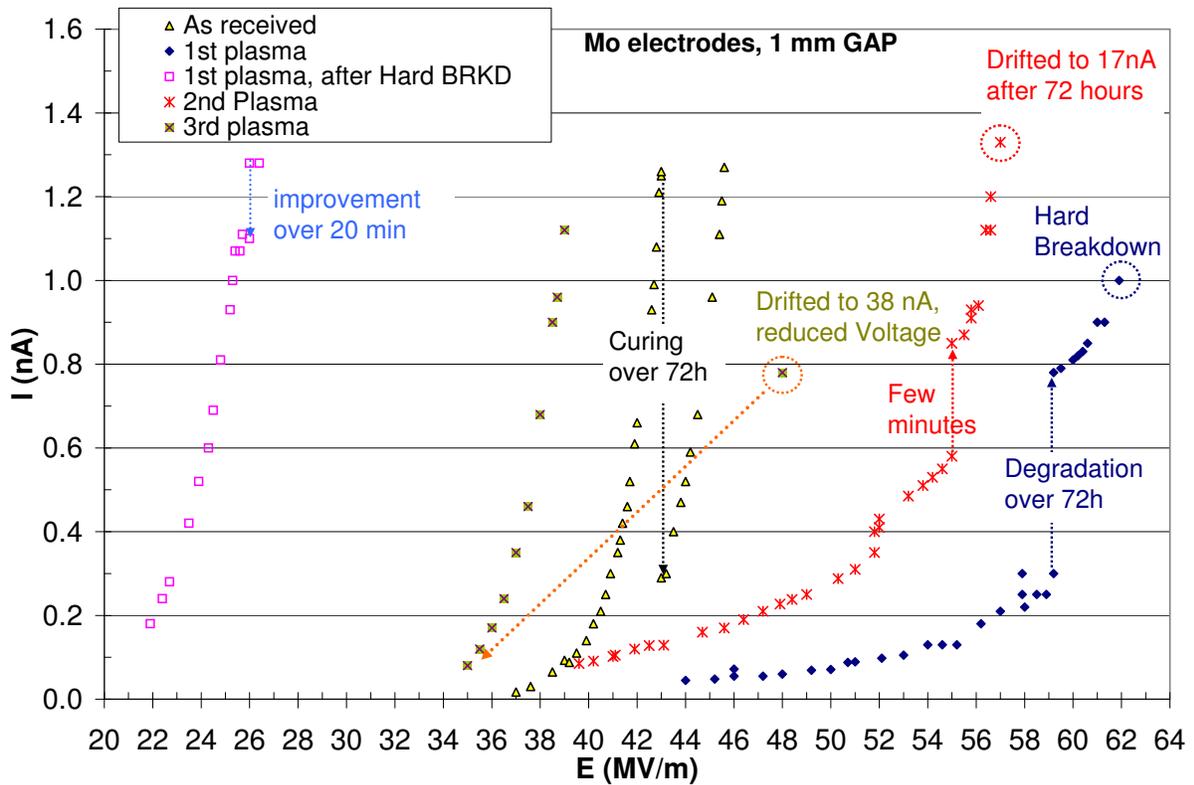}
\caption{Dark current evolution of Mo electrodes after
processing.} \label{figMochart}
\end{figure}

As for the Ti electrodes, dark current appears at some voltage,
contrary to Al, Cu and SS electrodes where dark current appears
only after a breakdown. Raising the field further increased the
dark current until breakdown occurred. Dark current plots from as
received electrodes(triangles), and after three plasma treatments
are presented in Fig.\ref{figMochart}. The first ArGD pushed the
limits of the dark current onset (diamonds) until a hard breakdown
occurred, bringing the current into the $\mu$A range. The onset
and the 1~nA limit was then greatly reduced (squares). The second
ArGD enabled partial recuperation from this breakdown (asterisks).
A third plasma He-Ar, did not bring any extra improvement (crosses
in squares). Mo conditioning from broad electrodes at large gap
($> 500 \mu$m) was similar to conditioning with short gaps. Mo
reached a higher gradient than Cu but it sparked more to get there
\cite{Kildemo:2004}, hence showing more damage than Cu,
Fig.\ref{figCuMocathode}. From this, it is obvious that a special
electrode conditioning procedure must be used to process the Mo
electrodes, and thus to protect the electron source (FEA).

\subsubsection{Titanium vacuum fired}

Original Ti electrodes were re-installed after vacuum firing and
the results, for comparison with non fired Ti, are presented in
Table.\ref{tabTiSSMorslt}. After vacuum firing the Ti became gray
black. This color can indicate TiH$_2$ or TiO$_2$ in the rutile
form, or Ti$_3$O$_5$. Damage, all localized on the top of Ti
anode, can be seen in Fig.\ref{figTiMoCu}. The involuntary coating
is removed by the severe breakdowns sustained by both electrodes.

\subsubsection{Surface analysis} \label{xpsanalysis}
Given the rather poor performances in field holding for the Mo
electrodes, compared to literature values, and the colour of the
Ti electrodes, it is suspected that the vacuum firing degraded the
Mo properties instead of improving them. The pressure in the
vacuum furnace was probably around 10$^{-5}$~Torr. Ti and Mo
cathodes were sent for surface analysis, X-ray photoelectron (XPS)
and Auger (AES) spectroscopy. Upon AES analysis, the surfaces of
both electrodes exhibited high contents of carbon (C) and oxygen
(O), higher than normal air exposure. Nitrogen (N) is also present
on the Mo surface. Nitrogen cannot be seen on Ti as Ti and N
overlap under AES analysis. Under XPS analysis, a shift of about
5~eV to higher binding energy, for both Mo 3d$_{5/2}$ and Ti
2p$_{3/2}$ is observed. Such shift can be the mark of TiO$_2$ and
MoO$_3$ \cite{XPShandbook:92}. The Mo shows no discoloration,
suggesting the formation of a thin trioxide film. The conclusions
are, first of all a small air leak might have been present in the
furnace. Secondly, a bad heat treatment will bring the opposite
results in terms of field holding.

\section{Conclusions}

Plasma glow discharge is a very effective way to enhance the DC
field holding in between two broad electrodes. It also permits
restoration and sometimes improvement of the DC field limit
achieved after a breakdown event leading to dark current emission.
This dark current follows the FN law.

The downside of such treatment, for Al, Cu and SS electrodes, is
that the surfaces hold the field until breakdown with no or little
warning. No increase in pressure is recorded, but sometime some
current variation in the tens of pA might appear. Such
fluctuations can be the sign of a forthcoming breakdown, if any,
but the time scale can vary from minutes to hours. In the hunt for
the breakdown precursor, in the framework of an interlock for the
500~kV pulser, a highly sensitive and fast photomultiplier will be
tested in this 100~kV DC test stand.

For Ti and Mo electrodes dark current appears and increases when
raising the voltage until breakdown occurs. However, during the
processing a few sparks occur, sometimes in a "spitfest" regime.
Those sparks are beneficial as they condition the surface. Dark
current at a level of 1~nA can either drift to hundreds of nA,
fall back to less than a hundred pA or stay stable. So far, the
prediction for its evolution is only empirical.

Finally we have, without stringent procedures, matched or exceeded
results obtained by other labs. However, results for non mirror
finished Ti and Mo were below the ones obtained elsewhere. In the
case of Mo, it is suspected that the vacuum firing contaminated
the Mo as it did for Ti, leading to poorer performances than
usually reported in the literature.

In order to find the Grail material, which will hold our requested
field without emitting dark current, Niobium seems to be a
material of interest. Ion implantation, with nitrogen, is known to
harden materials \cite{Woolley:1997,Shokouhy:ICPIG05}. It may be
possible that this technique of hardening can be useful to
increase the breakdown threshold of soft materials, as it seems to
have for harder ones \cite{Sinclair:PAC01}. Dark current from
electrodes can be lowered by depositing a pure monolayer of oxygen
on the surface, which will increase the work function of the
electrodes. However, and in the framework of an accelerator
electron source, this layer might have to be regenerated
frequently as back bombardement from residual gas ions will clean
the surface.

\section{Acknowledgments}

Kugler GMBH for the donation of the mirror finished Al and Cu
cathodes. R. Kirby (SLAC) and U. Mueller (EMPA Duebendorf) for the
surface analysis. S. Ritter (PSI) for the SEM time. Finally, to M.
Taborelli, T. Ramsvik, S. Sgobba at CERN and E. Kirk (PSI) for
some useful discussion.

%
%

\clearpage

\begin{table}[htbp]
\begin{center}
\caption{Secondary electron yield maximum \cite{crc}, sputtering
yield by 500~eV incident Ar \cite{Vossen:78} and self-sputtering
rate at 500~eV of different elements
\cite{Posadowski:1993,Anders:1995}.}
\begin{tabular}{|c|c|c|c|c|c|}
\hline Elements& SEYmax& Atm/Ar inc& Self Sputter & Melting Point & Young Modulus \\
 & & & rate& T$^\circ$C  & GPa \\
 \hline Cu& 1.3& 2.3 & $>\quad 1$&
1083& 110 \\
\hline Al& 1.0& 1.05 & $<\quad 1$&
660 & 69 \\
\hline Au& 1.4& 2.4& $>\quad 1$&
1063 & 78 \\
\hline Ti& 0.9& 0.5 & $<\quad 1$&
1668 & 116\\
\hline Mo& 1.25& 0.6 & $<\quad 1$&
2610 & 329\\
\hline Zr& 1.1& 0.65 & $<\quad 1$&
1852 & 68\\
\hline Fe& 1.3& 1 (SS 1.3)& $\sim\quad 1$ \cite{Mason:94}&
1536 & 200 \\
\hline W& 1.4& 0.57 & $<\quad 1$ &
3410 & 411\\
\hline Ta& 1.3& 0.57 & $<\quad 1$ &
2996 & 186\\
\hline Nb& 1.2& 0.6 & $<\quad 1$ &
2415 & 105\\
\hline
\end{tabular}
\label{tabElectrchoice}
\end{center}
\end{table}

\begin{table}[htbp]
\centering \caption{Measurement history of air-exposed dark
current electrodes.}
\begin{tabular}{|c|c|c|c|c|}
\hline \raisebox{-1.50ex}[0.4cm][0cm]{\textbf{Cathode}}&
\raisebox{-1.50ex}[0.4cm][0cm]{\textbf{Anode}}& \textbf{As}&
\textbf{Plasma}& \raisebox{-1.50ex}[0.4cm][0cm]{\textbf{n$^{th}$ Plasma}} \\
& & \textbf{received}& \textbf{(He - Ar)}&  \\
\hline SS& SS& Yes& Yes& Yes \\
\hline Al& Al& Yes& Yes& Yes \\
\hline Al mirror Finished& Al (sme as abv)& Yes& Yes& Yes \\
\hline Cu oxidized& Cu oxidized& -& Yes& Yes \\
\hline Cu Polynox{\texttrademark}& Cu Polynox{\texttrademark}&
Yes& Yes& - \\
\hline Ti& Ti& Yes& Yes& Yes \\
\hline Mo vac fired& Mo vac fired& Yes & Solely Ar & Yes \\
\hline Ti vac fired & Ti vac fired& Yes & Solely Ar & Yes \\
\hline Cu mirror Finished& Mo vac fired& Yes& Solely Ar & Yes \\
\hline
\end{tabular}
\label{tabHistoryelectrode}
\end{table}

\begin{table}[htbp]
\centering \caption{Field gradient (MV/m) between electrodes
obtained at 1~mm gap for 1~nA of dark current or with no field
emission (FE), second row.}
\begin{tabular}{|c|c|c|c|c|c|c|c|}
\hline  & SUS& Cu& Ti& Mo&
Mo - Ti & Al & Nb\\
\hline 1 nA \cite{Furuta:Nima2005}& 36& 47.5& 88& 84&
103 & - & -\\
\hline No FE \cite{diamond1:98}& - & 70 & 60 & -&
- & 85 & 92 \\
\hline
\end{tabular}
\label{tabFuruta}
\end{table}

\begin{table}[h]
\begin{center}
\caption{Electric field in MV/m held in between two Al electrodes
at 1~mm gap for the given dark current in nA.}
\begin{tabular}{|c|c|c|c|c|}
\hline & \multicolumn{2}c{Al - Al} & \multicolumn{2}{|c|}{Al mirror finished - Al}\\
\hline State / Dark Current & $<\quad$0.05~nA& 1 nA& $<\quad$0.05~nA& 1 nA \\
\hline As Received & - & 7.5 & 36 (2 mm)& 29\\
\hline After Plasma & 52 & 30 & 73 (stable) &31 \\
& & & (92 at 750$\mu $m)& \\
\hline
\end{tabular}
\label{tabAlRslt}
\end{center}
\end{table}

\begin{table}[htbp]
\begin{center}
\caption{Electric field in MV/m held in between Cu electrodes at
1~mm gap for the given dark current in nA. (*) obtained at 3mm
gap}
\begin{tabular}{|c|c|c|c|c|c|c|}
\hline & \multicolumn{2}c{Cu oxidized} & \multicolumn{2}{|c|}{Cu clean}& \multicolumn{2}{c|}{Cu-Mo} \\
\hline State / Dark Current & $<\quad$0.05~nA& 1 nA& $<\quad$0.05~nA& 1 nA & $<\quad$0.05~nA& 1 nA \\
\hline As Received& - & -& 24& 26 & 18.2 (*)& 13.8 (*)\\
\hline After Plasma& 32 & 29.3 & 55 & 19 & 21.6 & 25.4\\
\hline
\end{tabular}
\label{tabCuRslt}
\end{center}
\end{table}

\begin{table}[htbp]
\begin{center}
\caption{Electric field in MV/m held in between two SS two Ti and
two Mo electrodes at 1~mm gap for the given dark current in nA.}
\begin{tabular}{|c|c|c|c|}
\hline & State / Dark Current& $<\quad 0.05$ nA&1 nA \\
\hline \raisebox{-1.50ex}[0cm][0cm]{SS}& As Received& 40&42.5 \\
\cline{2-4} & After Plasma& 68 & 35 \\
\hline \raisebox{-1.50ex}[0cm][0cm]{Ti}& As Received& 50&46.6 \\
\cline{2-4} & After Plasma& 63 & 67 (0.1nA) \\
\hline \raisebox{-1.0ex}[0cm][0.3cm]{Ti}& As Received &  29.6 & 32.5 \\
\cline{2-4} Vac Fired & After Plasma& 39 & 41.4 \\
\hline \raisebox{-1.0ex}[0cm][0.3cm]{Mo}& As Received& 37 & 45.2 \\
\cline{2-4} Vac Fired & After Plasma& 44 &61.3 \\
\hline
\end{tabular}
\label{tabTiSSMorslt}
\end{center}
\end{table}

\clearpage

\clearpage

\clearpage

\clearpage

\listoftables
\newpage
\listoffigures

\end{document}